# The Normalization Problem of Special Relativity

Franz–Günter Winkler[1]

Abstract

*For the special theory of relativity, the normalization problem is formulated as the question how observers in constant relative motion may reach an agreement on space and time scales. As the normalization problem does not receive a thorough treatment in the standard development of the theory, two possible solutions called normalization at rest and normalization in motion are suggested. It is shown that normalization in motion makes the validity of the Lorentz transformation the result of mere conventions. The non-conventional approach following from normalization at rest requires an assumption concerning the acceleration of objects. A conceptual distinction between inside and outside observation, together with the analysis of the normalization problem, allows us to formulate an alternative interpretation of special relativity contrasting both the standard interpretation and Lorentz-type interpretations.*

## 1. INTRODUCTION

In this paper, an investigation into the meaning of special relativity is attempted, which starts from the question how observers in relative motion may achieve an agreement on their space and time scales. As this *normalization problem* does not receive a thorough treatment in Einstein's development of the Lorentz transformations, two possible solutions are discussed in the framework of special relativity. They will be called *normalization at rest* and *normalization in motion*. It will be shown that these solutions are related to different interpretations of the Lorentz transformations. In the case of normalization in motion, the Lorentz transformations can be developed by the exclusive use of conventions; in fact, the relativity principle and the principle of the invariance of the speed of light would play the role of mere conventions for a solution to the normalization problem. Normalization at rest requires an assumption concerning the acceleration of objects. It is shown that an adequate assumption can replace the relativity principle in the derivation of the Lorentz transformations.

Neither normalization in motion nor normalization at rest seems to be compatible with the standard interpretation of special relativity. Therefore, an alternative interpretation of special relativity not only deviating from the standard interpretation, but also from Lorentz-type interpretations is suggested. It is based on the analysis of the normalization problem and on a conceptual distinction between *inside* and *outside* observation.

## 2. THE NORMALIZATION PROBLEM

To measure means to compare. In hard science, the conditions under which a measurement takes place have to be defined very strictly. Most importantly, it has to be assured that the measuring instruments, i.e. the tools with which the investigated phenomena are compared, are in a well-defined sense the *same* for every measurement. Only when this is guaranteed to a certain extent, it is justified to relate the measured phenomena to each other.

Out of these basic considerations about the measurement procedure a question arises for the special theory of relativity, which deals with measurements stemming from observers in relative motion: By which procedure do different observers (inertial frames of reference) assure that their measuring instruments are the same, and, consequently, in which sense can space and time scales held by different observers be considered the same?

---

[1] Franz-Günter Winkler, Stumpergasse 63/11, A-1060 Vienna, Austria. fg.winkler@aon.at

This question will be analyzed as the *normalization problem of special relativity.* In Einstein's development of the relativistic calculus, the normalization problem is not addressed. It is, though, a problem that has to be solved whenever measurements stemming from different observers are related to each other as is done in the Lorentz transformations.

In order to make the role of the normalization problem for the special theory of relativity and the implications for its interpretation explicit, two possible solutions to the normalization problem will be discussed, which will be called *normalization at rest* and *normalization in motion*.

## 2.1 Normalization at Rest

The most natural way to solve the normalization problem is to build two identical sets of measuring instruments (meter sticks and clocks) that rest in a first frame of reference and transfer one of them to a second, moving frame. After that the measuring instruments of the second frame can be adapted to the first frame. This procedure requires the acceleration of a set of measuring instruments.

Though avoiding the term acceleration, Einstein seems to have normalization at rest in mind when introducing the relativity of lengths.[1] In his scenario, only one meter stick is used, yet in two different (constant) states of motion. At first, the meter stick and some object rest in frame 1. The length of the object is measured by the meter stick. Then the same meter stick and the same object appear as resting in another frame 2 showing some constant motion relative to frame 1. Again, a measurement of the object takes place by the use of the meter stick. According to the relativity principle, it is postulated that both measurements must yield the same value for the length of the object. As a consequence, it turns out that both meter stick and object resting in frame 2 show a contraction when being measured in frame 1 (and vice versa).

In the case of normalization at rest, the space and time scales of different frames are adapted with the help of an accelerated set of measuring instruments. By this, the identity of the scales of different frames is directly given. In the case of normalization in motion, such direct adaptation is impossible.

## 2.2 Normalization in Motion

As normalization at rest is linked to the acceleration of objects, it should not be the choice of the proponents of the standard interpretation, who regard length contraction and time dilation as properties of space-time as such. As an adequate basis of this view, a different class of normalization procedures can be introduced, which will be called *normalization in motion*. The basic idea of normalization in motion is that two observers aiming at an agreement on meters and seconds stay in constant relative motion when performing the normalization procedure.

In the following, three versions of normalization in motion are offered, all of which are based on the relativity principle. Two observers solving their normalization problem in relative motion choose their scales of space and time such that the relativity principle is satisfied.

**2.2.1 Normalization by Postulated Identity (I).** It is usually taken as an empirical fact that certain physical objects or processes have identical spatial or temporal extensions from the view of their rest frames. Good examples are decay processes of atoms producing radiation of some fixed wavelength. Yet, as long as observers in different states of motion performing an experiment have not come to an agreement on meters and seconds, this extensional identity can only be a postulate. This postulate, however, could be interpreted as an instruction for different observers to solve their normalization problem, namely in the following sense: Every observer has to choose the length of his meter stick such that the selected process has the postulated extension. By this, the relativity principle is satisfied, as the measured extensions depend exclusively on the relative speed with respect to the observer. (The relative speed equals zero in this trivial case.)

**2.2.2 Normalization by the Use of a Moving Object (II).** As will be shown in section 3, the normalization problem need not be solved in order to allow different observers to perform useful and compatible measurements of velocities. On this basis, any two non-normalized observers in relative constant motion may arrange a scenario, in which an object moves with some velocity *v* from the view of the first observer and with the velocity –*v* from the view of the second observer. According to the relativity principle, both observers should measure the length of the object to the same value. Again, two observers could take this assertion as a basis for the normalization of their meter sticks.

**2.2.3 Normalization by mutual measurement (III).** The minimal construction involves only two observers agreeing on their relative motion. Other than procedure (II), which requires a referential object resting in a third frame, procedure (III) is based on mutual measurements of meter sticks resting in the frames that are to be adapted. The involved observers vary the lengths of their (arbitrarily chosen) meter sticks until they measure each other's meter sticks to the same value. This solution, like the previous one, presupposes compatible measurements of velocities by non-normalized observers.

Though all three solutions to the normalization problem are built upon the relativity principle, they are not equivalent from a philosophical point of view. Solution (I) requires the assumption that there are copies of physical objects or processes which can be treated the same in different states of motion by co-moving observers. Solutions (II) and (III) do not make use of copies. The clarification of this difference requires a closer look at the relativity principle.

The relativity principle postulates that the *form* of the laws of physics is the same in all inertial frames. Using the relativity principle as a basis for a solution to the normalization problem - to be precise - goes beyond that meaning, insofar as the form of a law has nothing to do with the scaling of the measured variables.

How can two observers not yet having solved their normalization problem conclude that they have to do with the same objects or processes when performing some experiment in their rest frames? All they can do is to compare the set of relations between the variables they measured in their own experiment with the set of relations between the variables measured by the other observer in his experiment. If the relations are equivalent, then it is justified to assume that the same *types* of objects or processes were involved in both experiments. However, it is not legitimate to conclude that also the extensions must be identical. Formal equivalence does not imply equivalence with respect to scaling. Only by an act of definition, i.e. by setting the scales accordingly (normalization procedure (I)), are the extensions of the two objects or processes made identical.

Normalization in relative motion, irrespective of the chosen procedure, is a legitimate option, as the relativity principle is available in the axiomatic system leading to the Lorentz transformations. However, the implications for the interpretation of the Lorentz transformations are enormous. If the normalization problem is solved at relative rest by the use of accelerated measuring instruments, the validity of the Lorentz transformations is an empirical question: Is it true that an accelerated meter stick shows the expected contraction? If, instead, the relativity principle comes in as a *convention* for the normalization of measuring instruments resting in inertial frames, the validity of the Lorentz transformations is tautological. This difference will be demonstrated in a thought experiment in section 6.

## 3. COORDINATE TRANSFORMATIONS FOR NON-NORMALIZED OBSERVERS

There are many different ways to derive the Lorentz transformations (e.g. [2, 3]), all of which ignore the normalization problem. In the following, a development of the Lorentz transformations is suggested that makes the treatment of the normalization problem explicit. As a first step, a set of coordinate transformations is introduced that leaves the normalization problem still open. In section 4 it is shown how normalization in motion leads to the Lorentz transformations, in section 5 the same is shown for normalization at rest.

## 3.1 Space-Time Frames

The presented approach to special relativity makes use of a fully Euclidean space-time geometry. The central concept of a space-time frame and its relation to measurements of space and time distances can be characterized as follows.

**3.1.1 Measurements and space-time frames.** On the one hand a space-time frame has to be regarded as a geometrical construction on the basis of measurements of space and time distances performed by an observer, on the other hand every space-time frame can be used to describe the measurement processes performed by observers in relative motion. It is thus possible for every observer holding a frame to reconstruct the frames held by observers in relative motion by measuring their measuring processes.

As a consequence of this, special relativity can be developed from the perspective of any single frame. The equivalence of all inertial frames is a logical consequence of the validity of the Lorentz transformations, which can be derived in the Euclidean space-time geometry of a single observer. Before this is done, some remarks on the following argumentation have to be made.

For the purpose of this paper, it is sufficient to deal with only one space dimension in addition to the time dimension.[2] The orthogonality of the space and time axes is nothing more than a convenient choice making geometrical relations and calculations easy. There is no physical argument forcing the mapping of pairs of measured space and time distances between events to a rectangular geometry. In the diagrams, the time axis is scaled with the constant *c* (speed of light), which makes light rays appear as 45° lines.

---

[2] Thus, the normalization problem is considered only for meter sticks showing the same spatial orientation. A more complete coverage of the normalization problem would comprise the adaptation of meter sticks with different spatial orientations. Similar to the normalization problem for moving meter sticks in one dimension, there is a *conventional option*, which is based on the conventional constancy of the speed of light in all directions, and a *non-conventional option*, which is based on the rotation of a meter stick.

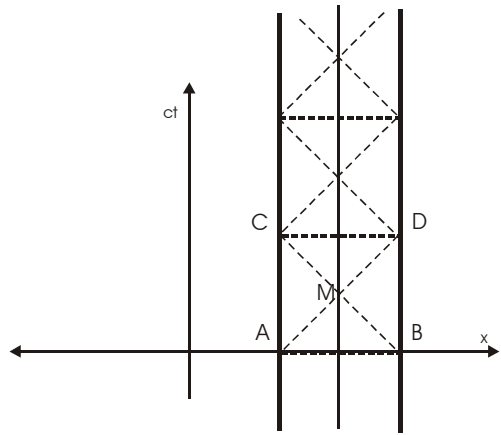

Fig. 1. A resting light clock consisting of two mirrors between which two light rays are reflected. The light rays meet in the middle point M. Events A and B respectively C and D are simultaneous.

In figure 1 the concept of a light clock is introduced in the framework of an observer, for whom the invariance of the speed of light is assumed. In addition to Einstein's light clock a second light ray is reflected between two mirrors, which are fastened at the end points of a stick. The light rays are synchronized such that they always meet in the middle point *M* of the stick. By this, Einstein's definition of synchrony holds for events *A* and *B* as well as for events *C* and *D*.

While figure 1 shows a resting light clock, the light clock of figure 2 is in motion along the x-axis. For the moving clock, the events *A'* and *B'* as well as the events *C'* and *D'* are synchronous. The associated space and time axes of the moving clock show an angle $\gamma$ relative to the axes of the rest frame.

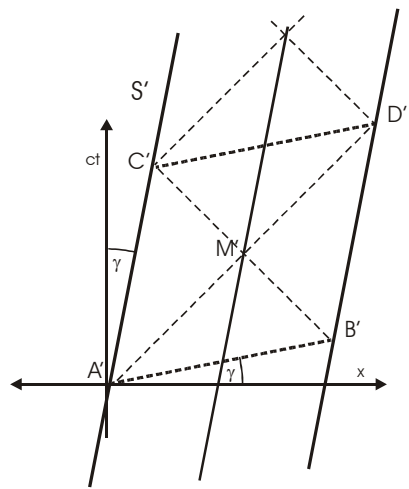

Fig. 2. A moving light clock. Events A' and B' respectively events C' and D' are simultaneous.

The clocks in figure 1 and figure 2 represent inertial observers holding their own space-time frames. The parallelograms *ABCD* respectively *A'B'C'D'* can be regarded as the elementary cells of these frames, which hold space-time coordinates for all possible events, including all events belonging to the measurement procedures performed by different observers.

The assumption that light has an invariant speed for both observers can very easily be implemented by the following convention.[3]

---

[3] Note that the assumption of the invariance of the speed of light (for all inertial observers) presupposes a solution to the normalization problem. As long as the normalization problem is not solved, the invariance of the speed of light cannot be an empirical fact, but only a convention.

**3.1.2 Conventional invariance of the speed of light.** In order to make the speed of light invariant in all inertial frames, each observer takes the arbitrarily chosen length of his light clock to be 1 meter and the temporal interval between two reflections ("ticks") as 1 second divided by $c$. If all observers do so, the speed of light is fixed to $c$. From now on, it is assumed that for all inertial observers – by this *convention* – the speed of light is $c$.

Before showing how frames can be mapped to each other, the mutual measurements of two observers are illustrated by the use of the light clocks that represent their frames. Figure 3 shows two observers (light clocks) in relative motion who take the length of their own stick to be *1 meter*. The simultaneous measurement of the other observer's stick marks a Euclidean space-time distance that is compared to the meter stick of the observer performing the measurement. For *S'* the length of *S* is *x'* divided by *1 meter(S')*. For *S* the length of *S'* is *x* divided by *1 meter(S)*.

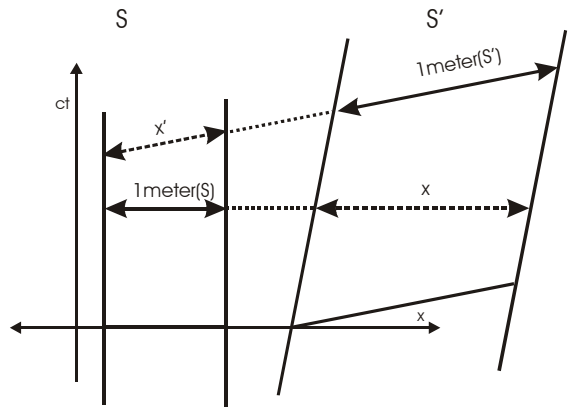

Fig. 3. Two observers measure each other's lengths simultaneously from their own point of view.

## 3.2 Derivation of the Transformations for Space and Time Distances

The transformations of space and time distances held by a resting frame *S* to a frame *S'* moving with velocity *v* can be derived using figure 4.

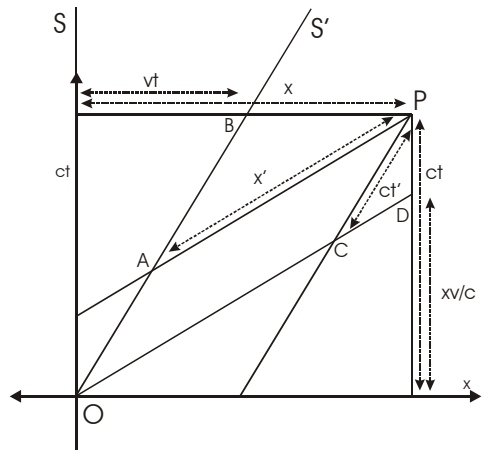

Fig. 4. The coordinates of event *P* are measured in the resting frame *S* and in the frame *S'* which moves with velocity *v* relative to *S*. Both coordinate systems are assumed to have the same origin *O*. The similarity of the triangles *ABP* and *CDP* allows to formulate the transformations for space and time distances from frame *S* to frame *S'*.

The two non-normalized frames in figure 4 have a common origin in event *O*. In both frames the space and time coordinates of event *P* are measured. The diagram shows the perspective of frame *S*, whose time axis is scaled with $c$. However, the following arguments hold for the perspective of any inertial observer. As has already been shown in figure 2, in such illustrations the space and time axes of frame *S'* have the same angle relative to the axis of the resting frame *S*. From this it follows that the triangles *ABP* and *CDP* are similar, which allows to formulate the following relation.

$$N = \frac{x'}{x - v \cdot t} = \frac{c \cdot t'}{c \cdot t - \frac{v \cdot x}{c}} \qquad (1)$$

The term *N* plays a key role for the normalization problem; therefore it will be called *normalization term*. As a consequence of (1), the transformations for space and time distances from frame *S* to frame *S'* take the form:

$$T(S => S')$$
$$\Delta x' = N \cdot (\Delta x - v \cdot \Delta t) \qquad (2)$$
$$\Delta t' = N \cdot (\Delta t - \frac{v \cdot \Delta x}{c^2})$$

The normalization term *N* is responsible for the scaling of the resulting space and time distances in *S'*. For each choice of the length of the meter stick in *S'* there is a corresponding value for *N*. It will be shown that the selection of a value for the normalization term *N* of a set of transformations *T (S=>S')* determines the value $N^{-1}$ of the inverse transformations $T^{-1}(S'=>S)$.

Before clarifying the relation between *N* and $N^{-1}$, three important properties of the transformations between in general not normalized observers have to be listed:

**(a) There is one signal whose speed is invariant for all observers.** This is already given by the way the space-time frames of inertial observers are constructed. For this construction, the invariance of the speed of light is a mere convention. In section 6, the assumption of an invariant signal other than light will be discussed.

**(b) Two observers agree on their relative velocity.** The validity of this statement can easily be demonstrated by describing the world line of a point resting in frame *S* and by transferring it to the frame *S'*, which moves relative to *S* with speed *v*.

$$\Delta x = 0 \cdot \Delta t$$
$$\Delta x' = N \cdot (0 - v \cdot \Delta t)$$
$$\Delta t' = N \cdot (\Delta t - \frac{v \cdot 0}{c^2}) \qquad (3)$$
$$v' = \frac{\Delta x'}{\Delta t'} = -v$$

Independent of the (non-zero) value of *N*, the speed of the frame *S* from the view of *S'* is –*v*.

**(c) The relativistic addition theorem for velocities holds.** The derivation of the addition theorem for velocities is essentially the same as when using the Lorentz transformations. In inertial frame *S* a point moves with constant speed $v_1$. The coordinates of that point are described from the view of a system *S'* that moves relative to *S* with speed $-v_2$. After that, the speed *v* of the point from the view of *S'* is calculated.

$$\Delta x = v_1 \cdot \Delta t$$
$$\Delta x' = N \cdot (\Delta x + v_2 \cdot \Delta t) = N \cdot (v_1 \cdot \Delta t + v_2 \cdot \Delta t)$$
$$\Delta t' = N \cdot (\Delta t + \frac{v_2 \cdot \Delta x}{c^2}) = N \cdot (\Delta t + \frac{v_1 \cdot v_2 \cdot \Delta t}{c^2}) \qquad (4)$$
$$v_1 \oplus v_2 = v = \frac{\Delta x'}{\Delta t'} = \frac{v_1 + v_2}{1 + \frac{v_1 \cdot v_2}{c^2}}$$

Also in this calculation, the value of *N* is irrelevant for the resulting velocity *v*. Once more it should be stressed that the statements (b) and (c) concerning velocities do not depend on a solution to the normalization problem. They also do not depend on the relativity principle. In the suggested Euclidean space-time geometry, they are derived from the definition of the measurement procedures (including the definition of synchrony) and the convention of the invariance of the speed of light.

## 3.3 The relation between two normalization terms

For the transformations *T(S=>S')* and for the inverse transformations *T⁻¹(S'=>S)* between two frames in relative motion, a change of the length of a meter stick in one frame always has an effect on both normalization terms. The relation between *N* and *N⁻¹* can be calculated by assuming that a space-time interval as being measured by frame *S* is mapped to itself when first being transformed to frame *S'* and afterwards being re-transformed to frame *S*.

$$T(S => S')$$
$$\Delta x' = N \cdot (\Delta x - v \cdot \Delta t)$$
$$\Delta t' = N \cdot (\Delta t - \frac{v \cdot \Delta x}{c^2})$$
$$T^{-1}(S' => S)$$
$$\Delta x = N^{-1} \cdot (\Delta x' + v \cdot \Delta t')$$
$$\Delta t = N^{-1} \cdot (\Delta t' + \frac{v \cdot \Delta x'}{c^2})$$

This leads straightforward to

$$N \cdot N^{-1} = \frac{1}{1 - \frac{v^2}{c^2}} \qquad (5)$$

As a consequence of both normalization in motion and normalization at rest, the values for *N* and *N⁻¹* will be shown to be equal, which leads to the Lorentz transformations. However, different and as well consistent[4] choices for *N* and *N⁻¹* are possible. E.g., a scenario can be imagined, in which all frames *S'* adapt their meter sticks such that the meter stick of a selected frame *S* is measured by them to 1 meter. Accordingly, the frames *S'* would measure also time intervals to the same values as frame *S*. Applying (2) and (5) leads to the normalization terms for this scenario.

$$N = \frac{1}{1 - \frac{v^2}{c^2}} \qquad N^{-1} = 1 \qquad (6)$$

## 4. NORMALIZATION BY THE RELATIVITY PRINCIPLE

The transformations *T* describe the mapping of space and time distances between inertial frames that have not solved their normalization problem, i.e. the lengths of the meter sticks have been chosen independently. The term *N* is responsible for the scaling of both space and time distances in *S'*. Procedures that fix the terms *N* and *N⁻¹* for the transformations between two frames and consequently the lengths of the observers' meter sticks are normalization procedures. One such normalization procedure follows from the relativity principle, saying: "Two inertial observers choose the lengths of their meter sticks such that they measure each other's meter stick to the same value." This normalization procedure has already been introduced as *normalization by mutual measurement* (III). In order to make the velocity of light equal *c* also after the normalization, the time scales have to be re-adapted according to section 3.1.2.

In the following calculation of the normalization terms for the transformations between two frames *S* and *S'*, the length of the meter stick of frame *S* as measured by frame *S'* is regarded as the result of the transformation of the space-time interval, which describes the length measurement of the meter stick of *S* performed by *S'*, from *S* to *S'*.

$$len'(meter(S))...\text{length of the meter stick of } S \text{ from the view of } S'$$

$$len'(meter(S)) = N \cdot (1 - \frac{v^2}{c^2}) \qquad (7)$$

---

[4] The consistency of such alternatives is limited to 1-dimensional space. The integration of a second or third spatial dimension is not compatible with the constancy of the speed of light for all spatial directions.

Applying (3.6) leads to

$$len'(meter(S)) = \frac{1}{N^{-1}} \tag{8}$$

The same procedure leads to the length of the meter stick of *S'* from the view of *S*.

$$len(meter(S')) = \frac{1}{N} \tag{9}$$

According to normalization procedure (III), the mutual measurements of the meter sticks must yield the same result.

$$len'(meter(S)) = len(meter(S')) \tag{10}$$

Therefore the two normalization terms *N* and *N⁻¹* must be identical, which allows to calculate their value.

$$N = N^{-1} = \frac{1}{\sqrt{1 - \frac{v^2}{c^2}}} \tag{11}$$

Inserting this expression for *N* to the formulae of the transformations *T* leads to the Lorentz transformations. This step completes a derivation of the Lorentz transformations which solves the normalization problem in relative motion by the use of the relativity principle.

However, this solution has a disadvantage which is connected to the interpretation of special relativity: Length contraction and time dilation are mere consequences of conventions, namely the constancy of the speed of light and the relativity principle, which stand behind normalization in motion. There is no possibility to conclude from this that an object which at first rests in one frame and then accelerates until it rests in the other frame shows the relativistic length contraction. Yet, this statement is part of the interpretation of special relativity as it stands.

In order to provide the full meaning of length contraction, an assumption on acceleration is inescapable and - as will be shown - sufficient.

## 5. NORMALIZATION BY AN ASSUMPTION ON ACCELERATION

Solving the normalization problem at rest requires an assumption on acceleration. In this section, an assumption is formulated which is motivated by Einstein's treatment of length contraction. The approach will be discussed in section 5.3.

## 5.1 Identity and Acceleration

The scenario in which Einstein introduces the contraction of lengths [1] consists of two measurements yielding the same value for the length of an object: At first, the object and the meter stick rest in some inertial frame, after that, both the meter stick and the measured object are in the same state of constant motion relative to the original rest frame. The identity of the measured lengths might be explained by a simple assumption on acceleration, which is not made explicit by Einstein, though.

**Hidden Assumption on Accelerating Objects.** Accelerating objects stay identical from their own perspective.

In order to make this assumption exploitable for mathematical analysis, the perspective of the accelerating object has to be defined. This can be done by the use of *tangent frames*.[5]

**Tangent Frame.** A tangent frame is an inertial frame from whose perspective a continuously accelerating object rests for a point in time.

In the infinitesimal limit, the velocities shortly before and shortly after this point in time take the same value for the tangent frame, differing only in direction.[6] It is therefore possible to regard the tangent frame as a kind of reflector for accelerating objects. Each continuous acceleration can be constructed from an infinity of infinitesimally small reflections or *"sudden changes"*.

---

[5] The concept of a tangent frame has been introduced by Einstein [4] and is common in general relativity.
[6] This holds whenever the second derivative of the object's world line exists.

Once the perspective of the accelerating object is defined as the perspective of the actual tangent frame, an assumption on acceleration can be formulated that supports Einstein's hidden assumption.

**Assumption of Simultaneous Change.** A change of the velocity of an object takes place simultaneously for the object.

On the basis of this assumption it will be shown in section 5.2 that any discrete sudden change of the velocity of an object leads to the Lorentz contraction and consequently to the relativistic normalization term *N*. The validity of this also for the limiting case of continuous acceleration is a mathematical consequence.

## 5.2 Derivation of the Normalization Term by the Assumption of Simultaneous Change

The assumption of simultaneous change allows to derive the normalization term *N* and thereby solves the normalization problem.

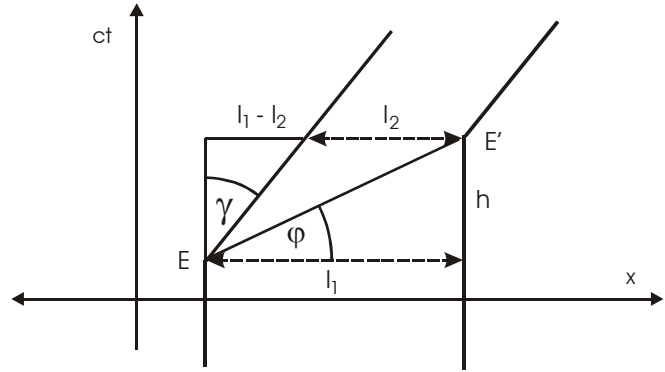

Fig. 5. The distance $l_1$ between two resting points changes to $l_2$, when they "suddenly" accelerate to the same velocity, but at different points in time. The events *E* and *E'* are simultaneous for the tangent frame, whose space axis shows an angle $\varphi$ relative to the resting frame's space axis. The time axis of the points after the sudden change show an angle $\gamma$ relative to the time axis of the resting frame.

Figure 5 shows the world lines of two resting points that suddenly change their velocity to *v*. They do so at events *E* and *E'* which are simultaneous from the view of some tangent frame moving with speed $v_T$. Equations (12) and (13) express the geometrical relations of the diagram.

$$\tan \gamma = \frac{v}{c} = \frac{l_1 - l_2}{h} \tag{12}$$

$$\tan \varphi = \frac{v_T}{c} = \frac{h}{l_1} \tag{13}$$

The fact that $v_T$ is the velocity of the tangent frame can be expressed by applying the addition theorem, which has already been derived for the transformations *T* between non-normalized inertial frames (4).

$$v = v_T \oplus v_T = \frac{2 \cdot v_T}{1 + \frac{v_T^2}{c^2}} \tag{14}$$

From these equations the contraction term *K* and consequently the normalization term $N^{-1}$ of the inverse transformations can be calculated.[7]

---

[7] The relation between the contraction term *K* and the normalization term $N^{-1}$ of the inverse transformations can be read from (8).

$$K = \frac{l_2}{l_1} = \sqrt{1 - \frac{v^2}{c^2}} \qquad (15)$$

$$N^{-1} = \frac{1}{K} = \frac{1}{\sqrt{1 - \frac{v^2}{c^2}}} \qquad (16)$$

Once the normalization term $N^{-1}$ of the inverse transformations takes this value, the normalization term $N$ must be identical according to (5).

$$N = \frac{1}{\sqrt{1 - \frac{v^2}{c^2}}} \qquad (17)$$

## 5.3 Discussion of the Assumption of Simultaneous Change

The assumption of simultaneous change leads to an accelerating behavior of extended physical objects which is known as *Fermi-Walker transport*.[5] It is connected with the concepts *rigid object* and *rigid motion of objects*.[6] Other than in the literature, where rigid motion is introduced by infinitesimal Lorentz transformations, it has been shown in this section that rigid motion, as an implementation of normalization at rest, allows a non-conventional derivation of the Lorentz transformations.

Building the Lorentz transformations upon the assumption of rigid objects creates a strange situation: Rigidity is not compatible with special relativity's limitation of physical effects to the speed of light. A rigid objects reacts "as a whole" when being accelerated, whereas from a classical, atomistic point of view such behavior is forbidden.

The non-rigid acceleration of physical objects produces tensions,[6] which means that spatial relations between parts of the object, as being measured by the tangent frame, undergo changes. However, this does not imply that the relativistic length relations do not hold after an object's acceleration. It could very well be the case that objects have the ability to restore the correct relativistic lengths after the tensions have come to an end. This would not only require that the object "memorizes" the spatial relations between its parts. It would also be necessary that the correct relativistic length is kept by at least one part, to which the lengths of all other parts could be re-adapted. As a consequence, rigidity has to be postulated at least for parts of physical objects in order to support normalization at rest and thereby the non-conventional meaning of length contraction.

As classical physics does not allow rigidity, it seems natural to seek rigidity and consequently a foundation of special relativity in the domain of quantum mechanics. The existence of non-local phenomena[7] as being postulated by quantum mechanics might provide an appropriate basis for this undertaking.

## 6. THE ROLE OF THE INVARIANT SIGNALING SPEED

The speed of light plays a key role for the conservation of an object's identity trough a process of acceleration: It is the simultaneity on the basis of the *speed of light* which brings forth the right contraction term of the Lorentz transformations. Thereby the definition of synchrony, the invariance of the speed of light, and the relativity principle loose their conventional character.

In the following, it will be shown that the conventional invariance of a speed different from that of light as the basis for the Lorentz transformations is in principle possible. The calculus works perfectly and does not lead to contradictions (e.g. time paradoxes). However, there are strong consequences for the description of acceleration and length contraction, as well as for scenarios involving signals that are faster than the chosen invariant signal.

## 6.1 Special Relativity in a Medium

As a thought experiment, the space-time frames of observers moving in a medium are analyzed who use the speed of sound waves in the medium ($c_l$) instead of light in the vacuum.[8]

Sound observers use sound clocks instead of light clocks. The space-time frames of sound observers are constructed in the same way as has been done for light observers: Constantly moving sound observers

---

[8] The discussion of a relativity theory based on a different invariant signal than light is inspired by Svozil [8].

- *choose rods of arbitrary lengths as their meter sticks and define the time between two ticks of their sound clocks as one second divided by $c_1$,*
- *adapt their meter sticks such that they measure each others meter sticks to the same value (i.e. perform normalization procedure (III)), and*
- *re-adapt their time scales such that the speed of sound, again, equals $c_1$.*

From the same arguments as in sections 3 and 4 it follows that the transformations between two frames take the form:[9]

$$\Delta x' = \frac{(\Delta x - v \cdot \Delta t)}{\sqrt{1 - \frac{v^2}{c_1^2}}} \qquad \Delta t' = \frac{(\Delta t - \frac{v \cdot \Delta x}{c_1^2})}{\sqrt{1 - \frac{v^2}{c_1^2}}} \qquad (18)$$

As a consequence of these formulae, the motion of an observer relative to the medium cannot be read from the space-time measurements based on sound. However, when the readings of co-moving clocks based on different signaling velocities (sound and light) are compared, a difference appears which depends on the motion relative to the medium.

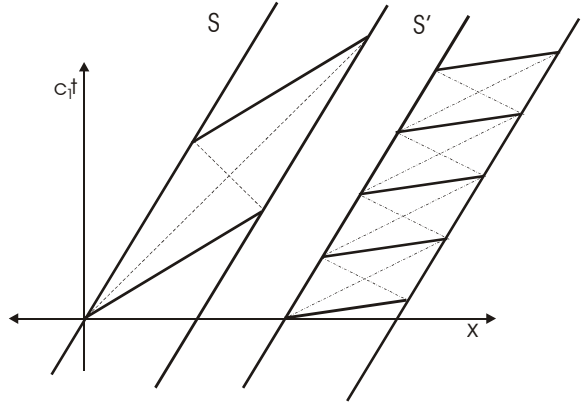

Fig. 6. From the rest frame of the medium, the space-time cells of a sound clock *C1* and a light clock *C2* moving at the same speed are illustrated. Observers using different clocks disagree on which events are simultaneous.

Figure 6 shows two such observers and their differing views of space and time. The observers no longer agree on which spatially separated events are simultaneous. From the view of the rest frame of the medium, the space axis of the sound observer is steeper than the space axis of the light observer. If, instead, both observers were at rest relative to the medium, their space axes would be identical.

More serious problems arise when objects are accelerated. The contraction terms of ordinary (light based) special relativity and of sound based special relativity result in different values. The fact that the sound based contraction term will not stand experimental tests shows the mere conventional character of sound relativity. Despite these shortcomings, there are no causal problems for sound relativity. Contrasting Svozil's approach,[7] it will be shown that the use of light signals does not lead to time paradoxes for sound observers.

## 6.2 Faster-than-Sound Communication

The following example shows the extreme case of faster-than-sound communication, namely when a signal is used that is "faster than the space axis" of a moving sound frame. In figure 7 two co-moving sound observers exchange light signals. Observer *O1* sends a light signal in the direction of observer *O2* at event *A*. The signal is received by *O2* at event *B'* and immediately answered by a signal directed at *O1*. This signal is received by *O1* at event *C*. For the observers *O1* and *O2*, the events *A* and *A'*, *B* and *B'* as well as *C* and *C'* are synchronous.

As can be read from the diagram, the two observers agree on the following statements.

---

[9] The relative velocity of the frames is limited to $c_1$.

- *The first signal moves backward in time and is received before its emission.   (A > B')*
- *The second signal moves forward in time and is received after its emission.   (C > B')*

Although it is possible to send signals "backward in time" in one direction, it is impossible to construct causal paradoxes: No event can ever depend causally on a later event at the same location $(A < C)$.[10]

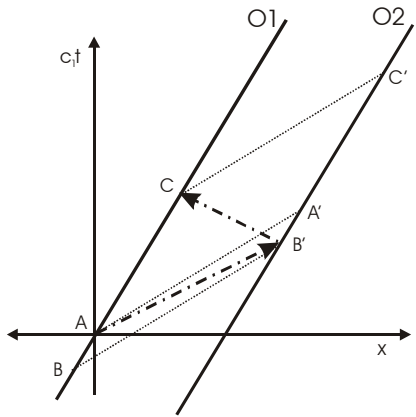

Fig. 7. Two sound-observers holding synchronized clocks are moving through the medium. Observer *O1* sends a light signal at event A to observer *O2*. The signal is received at event B' and immediately answered by another light signal which is received by observer *O1* at event C. In one direction, a light signal can destroy the temporal order between emission and reception (A > B), but not in both directions (B < C). Events taking place in one point in space cannot lose their temporal order (A < C). This means that causal paradoxes are not possible.

## 7. THE INTERPRETATION OF SPECIAL RELATIVITY

The analysis of the normalization problem results in a dilemma for the standard interpretation of special relativity: Either the validity of the Lorentz transformations is a product of mere conventions and therefore empirically meaningless, or an assumption concerning the acceleration of objects is required. In the following, an alternative interpretation of special relativity will be presented that not only contrasts the standard interpretation but also Lorentz-type interpretations.[3, 10, 11]

### 7.1 Standard Versus Lorentz-type Interpretations

The standard interpretation concentrates on Minkowski's geometrical picture, in which a unified *space-time* plays an ontological role. The invariance of the vacuum speed of light for any inertial observer is a property of space-time as such and is not due to any effects concerning the measuring instruments. Length contraction and time dilation are merely *perspective* – relative motion does not make a *real* difference between objects.

For defenders of a Lorentzian view, there is a twofold ontology concerning space and time. There exists an absolute frame, from whose perspective the substantial medium called *ether* is at rest, and an absolute flow of time. The invariance of the speed of light is valid only for the ether's rest frame and merely apparent for moving frames. Length contraction and time dilation, which depend on motion relative to the ether, are *real* phenomena with respect to the rest frame. The reciprocity of these effects, i.e. the fact that for the really contracted observer the resting observer also appears to be contracted, is due to the real contraction of the moving meter stick and to asynchronous measurements (according to Einstein's definition) performed by the moving observer.

As both interpretations make the same experimental predictions, the choice between them can only be based upon considerations external to the domain of special relativity. The main argument for adopting the standard interpretation is the fact that the assumption of an absolute rest frame is unnecessary for the development of the theory. Lorentz-type interpretations claim that they allow a classical understanding of special relativity by saving the medium for electromagnetic waves and the intuitive concept of an independent time dimension.

Although the two interpretations are significantly different, they can be attacked for the same weakness, namely for the secondary treatment of the normalization problem and consequently for ignoring the role of acceleration for special relativity.

---

[10] The example could be used in the discussion around hypothetical tachyons [9] Sound observers detecting light signals are in a similar situation as physicists would be when detecting faster-than-light signaling via tachyons.

## 7.2 An Alternative Interpretation

It has been argued that a clear treatment of the normalization problem makes it necessary to deal with acceleration. Otherwise the normalization of measuring instruments already presupposes the relativity principle and reduces to a mere convention leaving no empirical content for the Lorentz transformations. If, however, special relativity is built upon an assumption on acceleration, a different kind of interpretation of special relativity becomes possible.

## 7.2.2 How to Derive the Lorentz Transformations

The interpretation of a theory is closely linked to its derivation. According to the suggested interpretation the Lorentz transformations should be developed in a Euclidean space-time geometry by the following steps:

- *Definition of the measurement procedures according to Einstein.*

- *The assumption that the vacuum speed of light is invariant for one frame, independent of the speed of the source.*

- *The assumption that (ideal) objects accelerate simultaneously from their own point of view (which is equivalent to the assumption that accelerated objects show the Lorentz contraction).*

In this derivation, the relativity principle is missing. However, as a consequence of the Lorentz transformations the relativity principle gets a much stronger status. Unlike in the standard derivation, where the relativity principle boils down to a mere convention for the normalization of measuring instruments, it receives an empirical value in the alternative interpretation.

As has already been pointed out in section 5.4, the assumption of simultaneous change requires non-classical properties of matter. Therefore, quantum mechanics could become the basis for special relativity. For defenders of a Lorentzian view the possibility of simultaneous effects may seem attractive, as it might turn out that quantum mechanical effects are absolutely simultaneous, which would mark the rest frame of the ether.[11] This idea, however, is not supported by the proposed interpretation. The suggested simultaneity of acceleration depends on the motion state of the accelerating object and is not at all absolute. If acceleration was synchronized according to some absolute synchrony, the Lorentz contraction for accelerated objects would not hold in general.

## 7.2.3 Inside Versus Outside Observation

Both the standard interpretation and Lorentz-type interpretations make ontological claims. Before scheduling the alternative position in this question, a simple understanding of ontology is suggested which is based on a distinction between two types of observation.

In the final analysis, every physical observation is an *inside* operation in the sense that it is performed by an observer who is part of the universe. The measurement procedure is a physical process with two physical participants, namely the observer and the thing observed.[12] The result of every measurement has to be regarded as a product of a physical process involving both observer and thing observed and cannot be taken to capture a property of the observed thing as such.

From this fundamental argument it follows that a detached, *outside* observer position can only be a construction. This, of course, does not imply that dealing with outside conceptions is useless. It should be regarded as the very nature of human cognitive activity (including science) to construct outside models. Whatever else might be meant by ontology, it should contain the idea of properties that would be "visible" from a hypothetical outside observer position.

The alternative interpretation is aiming at a pair of inside and outside views fitting together in a way that makes clear how the outside view is *constructed* from the inside view and how the inside view can be *explained* from the outside view.[13]

The conceptual difference between the two types of observation can easily be applied to the measurements of space and time in special relativity. Thereby the distinction between inside and outside models is given a geometrical meaning. The outside view that is suggested as a basis for the alternative interpretation of special relativity is a 4-dimensional and fully Euclidean geometry. The diagrams presented in this paper are to be understood as 2-dimensional projections of this geometry. The outside view is comparable to the reader's view

---

[11] K. Popper[(12)] suggested to analyze the synchrony of quantum-mechanical effects as a possible test between standard and Lorentz interpretations of special relativity.

[12] More careful analysis, especially in the light of quantum theory, makes even the distinction between observer and thing observed appear as a result of an operation.[(13)]

[13] The distinction between inside and outside observation, which was introduced by the author more broadly in the context of cognitive science,[(14)] is in good accordance with Svozil's distinction between extrinsic and intrinsic observation.[(15)]

of the given illustrations. The reader may rotate the sheets of paper and check the diagrams from different sides. This, of course, does not relate to a physical process, because the outside observer is not part of what is being illustrated and does not undergo any changes (e.g. the reader's ruler stays the same).

Inside observers cannot be regarded as detached or unaffected when they change their space-time angle of observation, as their meter sticks and clocks do not stay the same. It is the inside observer transformation that is described by the Lorentz transformations. The Minkowski space-time geometry, which captures measurements stemming from inside observers, can *in principle* not take an ontological status in the above defined sense.

The 4-dimensional Euclidean world of the alternative interpretation also contrasts the twofold ontology of Lorentz-type interpretations consisting of a 3-dimensional space and a 1-dimensional time. From the constructed outside view there is one frame whose space and time axes appear as orthogonal lines, yet this frame should not be regarded as the absolute rest frame as is done by the Lorentz interpretation. Notions like space, time, motion or rest are – in the last consequence - meaningless for the outside observer and have to be understood as mere inside observer categories. The outside picture is completely static and leaves no room for an independent, dynamical time dimension.

From the alternative perspective, the key for a deeper understanding of relativity has to be sought in the nature of the physical object and its relation to the 4-dimensional whole. The existence of the physical object seems to be bound to two types of connections. On one hand the object maintains itself through time, on the other hand, as has been suggested, the spatially separated parts of the object must be coordinated simultaneously through space. The alternative interpretation - aiming at the construction of a consistent *outside* view - requires the analysis of these – from the *inside* - very different connections on a common ground.[14]

In some sense the alternative interpretation is more radical than both established interpretations. More explicitly than in the Lorentz interpretation, the understanding of special relativity is based on a constructed outside view. From this outside view, however, the integration of space and time into an inseparable whole reaches even further than in the standard interpretation, where space and time still receive a different treatment.

## 8. CONCLUSIONS

In the current physical debate, the question how to interpret special relativity is not taken very seriously. It is even hardly known that alternatives to the standard view in the form of Lorentz-type interpretations exist and that they are as well consistent. A major objective of this contribution is to underline that the interpretation of special relativity is an important and still open issue.

As an elementary argument about the normalization problem shows, there is a dilemma for the standard interpretation of special relativity. Either the Lorentz transformations are a mere product of conventions with no empirical content, or an assumption on acceleration is required, which deeply contradicts the spirit of the standard interpretation. The chosen assumption on acceleration can fully replace the relativity principle in the derivation of the Lorentz transformations. By explicitly presupposing non-local properties of matter, the presented approach suggests a foundation of special relativity in quantum mechanics.

The outlined interpretation of special relativity following from the treatment of the normalization problem contrasts both the standard interpretation and Lorentz-type interpretations. The differences are based on a distinction between inside and outside views. From the constructed outside view, the geometry of special relativity is 4-dimensional and fully Euclidean.

---

[14] The formulation of a conceptual basis for this has already been attempted by the author.[14,16]